\documentclass[a4paper,11pt]{article}
\usepackage{jinstpub} 
\usepackage{lineno}
\usepackage{needspace}

\title{\boldmath A New Concept of Liquid Xenon Time Projection Chamber for Medical Imaging}

\author[a]{Baron Li, }
\affiliation[a]{The Bishop's School, La Jolla, CA, USA}
\author[b]{Yue Ma, }
\author[b]{Kaixuan Ni}
\affiliation[b]{Department of Physics, University of California San Diego, La Jolla, CA, USA}

\emailAdd{nikx@ucsd.edu}

\abstract{
Liquid xenon time projection chambers offer a homogeneous detection medium with excellent intrinsic energy resolution, fast scintillation, and true three-dimensional position sensitivity, making them an attractive alternative to crystal-based detectors for positron emission tomography (PET). In this work, we present a new single-phase liquid xenon time projection chamber (TPC) concept optimized for medical imaging, employing combined scintillation and electroluminescence-based ionization readout to enable low-noise signal amplification and intrinsic depth-of-interaction measurement.

We evaluate the system-level performance of this detector concept using Monte Carlo simulations based on OpenGATE and Geant4, with direct comparison to conventional LYSO-based PET systems. The study focuses on detection sensitivity, energy-based event selection efficiency, and reconstructed spatial resolution. While LYSO detectors provide higher absolute stopping efficiency due to their higher density, liquid xenon detectors exhibit improved photopeak purity as a result of superior intrinsic energy resolution, leading to enhanced rejection of scattered events. 

Point-source reconstruction studies demonstrate that the intrinsic three-dimensional position sensitivity of the liquid xenon TPC translates into a reconstructed spatial resolution of approximately 1~mm full width at half maximum (FWHM) at the system level, compared to approximately 4~mm for LYSO-based systems under comparable conditions. These results indicate that liquid-xenon-based PET detectors can achieve competitive or superior imaging performance, particularly for applications requiring high spatial resolution, large axial acceptance, and scalable detector geometries.
}


\begin{document}
\maketitle

\section{Introduction and Motivation}

Positron emission tomography (PET) is a widely used medical imaging modality that enables quantitative visualization of metabolic and functional processes by detecting coincident pairs of 511~keV photons produced in positron--electron annihilation. Continued demand for higher sensitivity, improved spatial resolution, and extended axial coverage has driven the development of increasingly large and complex PET scanners. However, most commercial PET systems rely on segmented inorganic scintillator crystals, such as LYSO, coupled to silicon photomultipliers. While this technology provides excellent timing performance, it leads to high cost, fixed detector geometries, and limited intrinsic depth-of-interaction (DOI) capability~\cite{Trummer:2009zz, NIKNEJAD2017684}, ultimately limiting the spatial resolution that can be achieved in PET imaging~\cite{MOSES2011S236}.

Liquid xenon (LXe) offers an alternative detector medium with several intrinsic advantages for gamma-ray detection. Its high density and atomic number provide strong stopping power for 511~keV photons, while its large scintillation and ionization yields enable simultaneous optical and charge-based signal readout. LXe detectors also exhibit excellent intrinsic energy resolution and produce fast scintillation light. In addition, LXe is free of internal radioactivity and can be scaled to large, homogeneous detector volumes, offering flexibility in detector geometry and system design.

Earlier studies investigated liquid xenon (LXe) detectors for PET applications~\cite{Chepel:1999, DOKE2006863, ROMOLUQUE2020162397, Giboni_2007, Amaudruz_2009, Miceli_2011, Manzano:2018kxx}, but were limited by photosensor performance, cryogenic stability, and readout noise. Recent advances in VUV-sensitive photosensors~\cite{nEXO:2021uxc}, cryogenic purification~\cite{Plante:2022khm, Baudis:2023ywo}, and electroluminescence-based charge readout~\cite{Qi_2023, Martinez-Lema:2023zjk, Qi:2024naz} now enable a renewed evaluation of LXe-based PET detectors. In this work, we present a single-phase LXe time projection chamber concept optimized for medical imaging and evaluate its system-level sensitivity, efficiency, and reconstructed spatial resolution.
The detector design emphasizes high detection efficiency, intrinsic depth-of-interaction capability, competitive timing performance, and scalability to large detector volumes. Unlike conventional LYSO-based PET systems that rely on discrete crystal segmentation and limited DOI information, the LXe TPC provides a continuous detection medium with uniform response and true three-dimensional position sensitivity. %

\section{Detector Concept and Design Principles}

\begin{figure}[hbt!]
    \centering
    \includegraphics[width=0.9\linewidth]{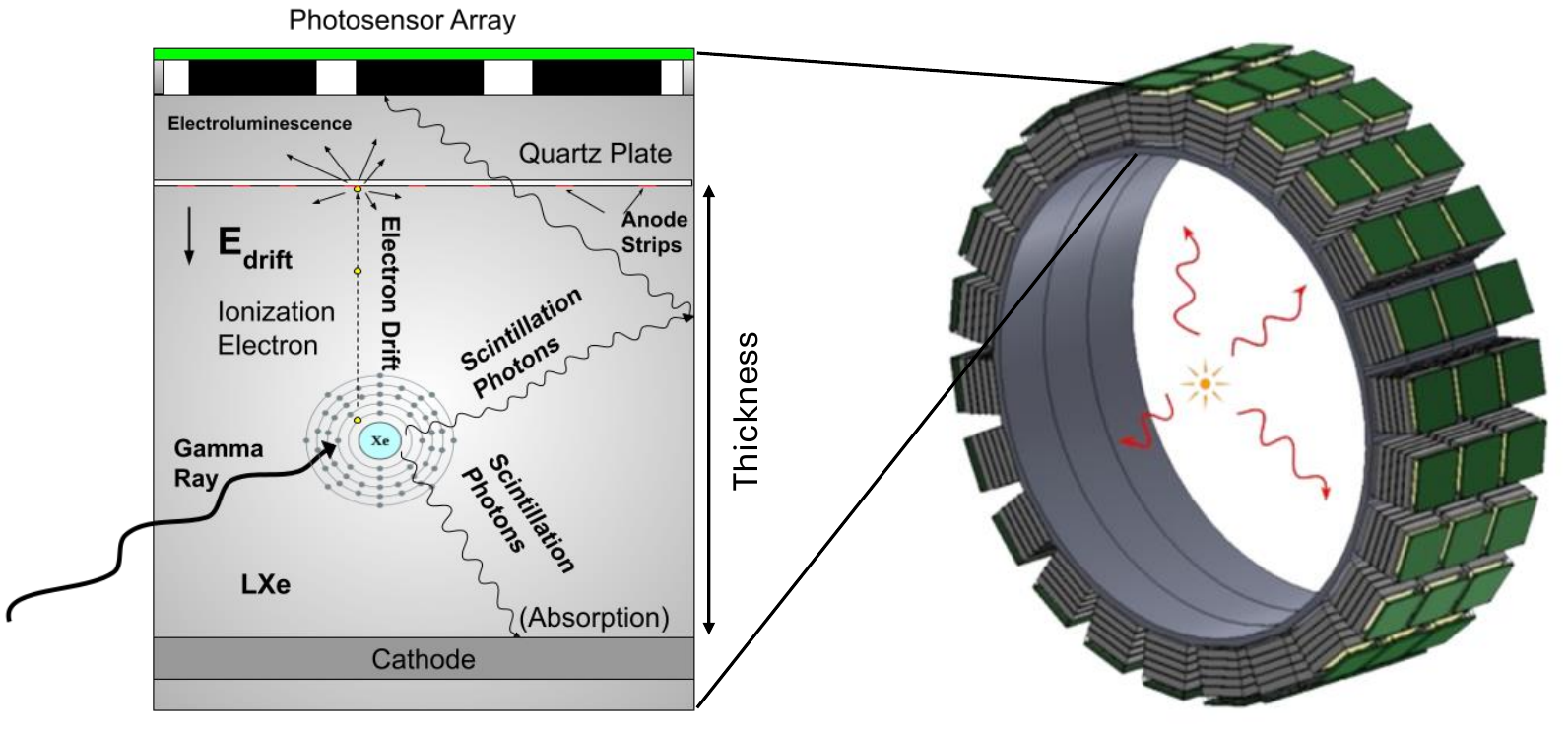}
    \caption{(Left) Operating principle of an individual liquid xenon (LXe) detector module, illustrating the detection of prompt scintillation light and delayed ionization signals via field-enhanced electroluminescence near the anode. The monolithic LXe volume provides high light collection efficiency and intrinsic depth-of-interaction sensitivity. Single-ended photosensor readout reduces channel count and system cost while maintaining sub-millimeter three-dimensional spatial resolution. The LXe target thickness can be adjusted to optimize detection efficiency and cost. (Right) Conceptual layout of a liquid-xenon-based PET imaging system consisting of concentric rings of LXe detector modules detecting coincident, back-to-back 511~keV gamma rays from positron–electron annihilation. Figures are adapted from \cite{Backues:2025}.}
\label{fig:lxepetdesign}
\end{figure}

The proposed detector concept is based on a single-phase liquid xenon time projection chamber operated at cryogenic temperature. Each detector module consists of a monolithic LXe volume instrumented with photosensors for prompt scintillation light detection and an electric field configuration that enables drift and readout of ionization electrons through field-enhanced electroluminescence (Fig.~\ref{fig:lxepetdesign}). Spatial information is obtained through continuous three-dimensional reconstruction rather than discrete crystal segmentation, providing depth-of-interaction information by design, without reliance on light-sharing or algorithmic inference. This architecture allows modular detector construction and scalability from compact organ-specific scanners to extended axial-length systems suitable for high-sensitivity imaging applications.

Gamma-ray interactions in liquid xenon produce prompt scintillation light and ionization electrons. The scintillation signal provides fast coincidence timing, while ionization electrons are drifted under an applied electric field to a localized electroluminescence (EL) region, where proportional light emission enables low-noise amplification and preserves the intrinsic energy resolution of liquid xenon. The time separation between scintillation and EL signals, together with the spatial distribution of the EL light, provides true three-dimensional position reconstruction and improved energy resolution through signal anti-correlation, enhancing scatter rejection. Implementing the EL region within or immediately adjacent to the liquid xenon volume avoids the need for a macroscopic liquid--gas interface, simplifying detector operation and improving mechanical stability. The resulting optical readout supports flexible segmentation and reduced electronic noise, enabling scalable detector geometries with improved spatial resolution and imaging uniformity, particularly for large-acceptance PET systems.

\section{Detector Performance: Sensitivity, Efficiency, and Spatial Resolution}

The intrinsic performance of individual LXe detector modules, including energy resolution and sub-millimeter position reconstruction, has been reported previously~\cite{Backues:2025}. In this section, we focus on system-level performance metrics relevant to PET imaging, including detection sensitivity, efficiency, and reconstructed spatial resolution.

\subsection{Simulation Framework}

PET scanner configurations employing LXe or LYSO detector modules were simulated using OpenGATE~10.0.0~\cite{Sarrut:2022ftg} package, built on Geant4~\cite{GEANT4:2002zbu}. The simulations modeled 511~keV gamma-ray interactions, energy deposition, and scintillation processes in detector modules of varying thicknesses. The simulated detector geometry consists of modules arranged in a circular ring with an inner radius of 354~mm and a radial thickness varied from 20~mm to 140~mm across the simulations. The axial width of the PET ring is approximately 164~mm. Housing, copper cooling plates, and thin shielding structures surrounding the ring were included and kept constant throughout the study. Simulation-level digitization models were implemented to incorporate realistic energy and position resolution, photon detection efficiency, and energy window selection.

\textcolor{red}{
}

\subsection{Detection Sensitivity and Efficiency}

The detection efficiency of a PET system depends on both the intrinsic stopping power of the detector material and the ability to identify and retain true full-energy events for image reconstruction. Using OpenGATE simulations, PET scanner configurations based on LXe and LYSO detector modules were compared over a range of detector thicknesses. The analysis focused on single-scatter, photo-absorbed 511 keV events. Simulation outputs consist of individual energy-deposition hits corresponding to primary gamma interactions in the detector medium (via photoabsorption or Compton scattering). Events were selected by requiring full energy absorption within a symmetric $2\sigma$ energy window determined by the detector energy resolution: $\sigma = 4.6$~keV for LXe (2.1\% FWHM)~\cite{Backues:2025} and $\sigma = 24.3$~keV for LYSO (11.2\% FWHM)~\cite{rausch2019philipsvereosperformance}. Multiple-scatter Compton events are naturally rejected in the LXe TPC due to their spatial separation, given the more than 1 cm mean free path of 511 keV Compton scattering.

\begin{figure}[hbt!]
    \centering
    \includegraphics[width=.49\linewidth]{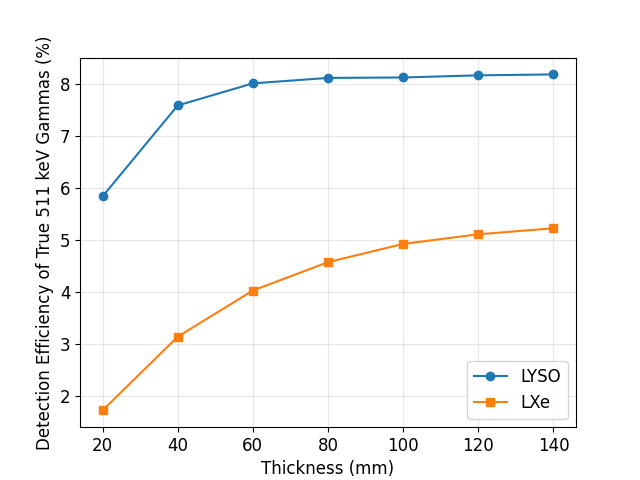}
    \includegraphics[width=.49\linewidth]{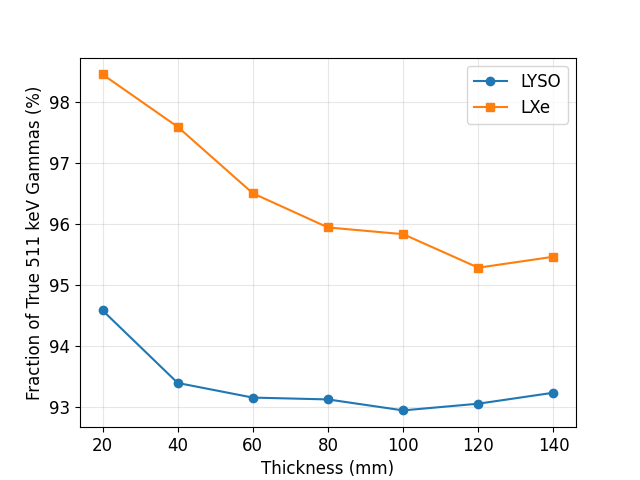}
    \caption{(Left) Detection efficiency of true 511~keV gamma rays within the $2\sigma$ energy window centered at 511~keV, relative to the total simulated events, for different radial thickness of the PET detectors. (Right) Fraction of true 511~keV gamma rays, relative to the total number of events, containing also scattered gammas, within the $2\sigma$ energy window centered at 511~keV. Statistical uncertainties (binomial) are at the level of ~0.005\% (left) and 0.02\% (right) and are smaller than the marker size; therefore, error bars are not shown.}
    \label{fig:efficiency}
\end{figure}

Owing to its higher density and effective atomic number, LYSO exhibits a higher absolute interaction probability for 511~keV gamma rays and therefore achieves higher raw detection efficiency across all simulated configurations (Fig.~\ref{fig:efficiency}, left). When restricting the analysis to true full-energy events, however, LXe detectors show a higher effective photopeak fraction, defined here as the fraction of true 511 keV interactions whose reconstructed energy falls within the selected photopeak energy window. The superior intrinsic energy resolution of LXe produces a narrower photopeak, improving discrimination between photoelectric absorption and Compton-scattered events and retaining a larger fraction of true 511 keV interactions within the energy window (Fig.~\ref{fig:efficiency}, right).

In addition, the monolithic LXe detection volume largely avoids the fine-scale inactive material and inter-crystal boundary losses associated with crystal segmentation and reflector gaps in conventional PET detectors, despite the presence of field cages and support structures in practical detector implementations. While the absolute stopping efficiency of LXe is lower than that of LYSO, the improved photopeak purity and reduced scatter contribution result in higher-quality coincidence datasets for image reconstruction, particularly in extended axial-length configurations where uniform detector response is critical.

\subsection{Reconstructed Spatial Resolution}

The reconstructed spatial resolution of the full PET system is a key system-level performance metric and provides a more meaningful comparison than intrinsic detector resolution alone. Spatial resolution was evaluated using point-source simulations reconstructed with the CASToR framework~\cite{Merlin_2018}, employing a list-mode ordered-subsets expectation maximization (OSEM) algorithm, with the source placed at the geometric center of the detector in both the radial and axial directions.

The reconstructed spatial resolution is governed by the intrinsic position resolution of the PET detectors. For LYSO-based systems, this resolution is determined by the crystal dimensions; standard 19~mm long crystals with a $4~\text{mm} \times 4~\text{mm}$ cross section were used in the simulations. For LXeTPC-based systems, intrinsic three-dimensional position sensitivity is provided by true depth-of-interaction information derived from the time separation between prompt scintillation and delayed electroluminescence signals, together with transverse position reconstruction from the electroluminescence light pattern. Based on electron diffusion in liquid xenon~\cite{NJOYA2020163965} and simulated reconstruction performance~\cite{Backues:2025}, position resolutions of 1~mm in depth and $2~\text{mm} \times 2~\text{mm}$ in the transverse plane were assumed.
In the LXe-based detector, this fine-grained interaction localization effectively corresponds to a large number of millimeter-scale pseudo detection elements, leading to a prohibitively large number of lines of response for full three-dimensional sinogram-based reconstruction. To illustrate the achievable system-level spatial resolution while maintaining computational tractability, only a single axial slice at the geometric center of the detector was reconstructed. The reconstructed point-spread function was quantified by fitting a two-dimensional Gaussian function to the reconstructed image after applying mild smoothing to suppress voxel-level fluctuations.

\begin{figure}[hbt!]
    \centering
    \includegraphics[width=.49\linewidth]{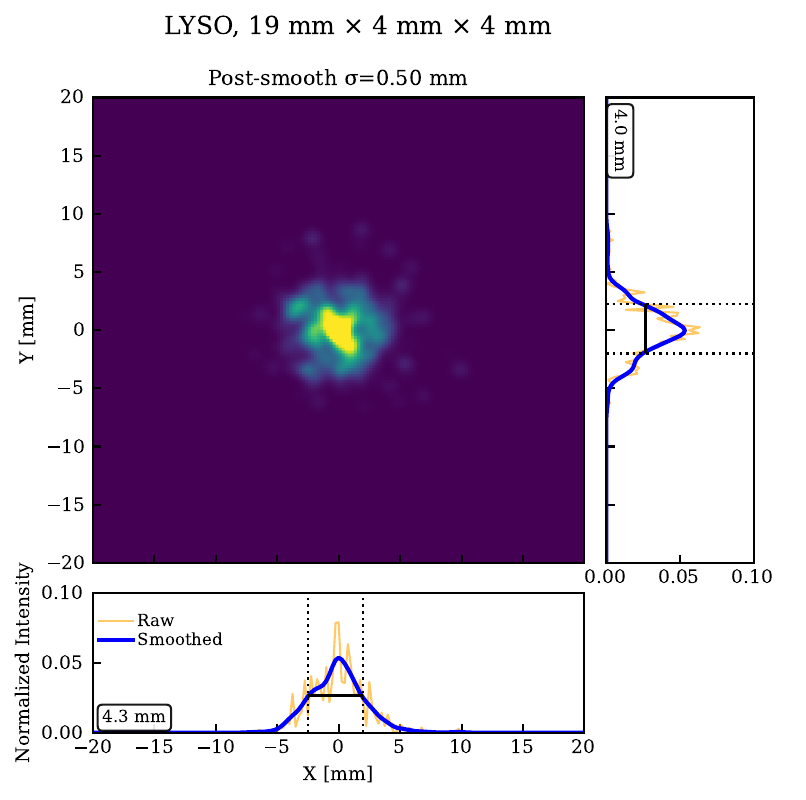}
    \includegraphics[width=.49\linewidth]{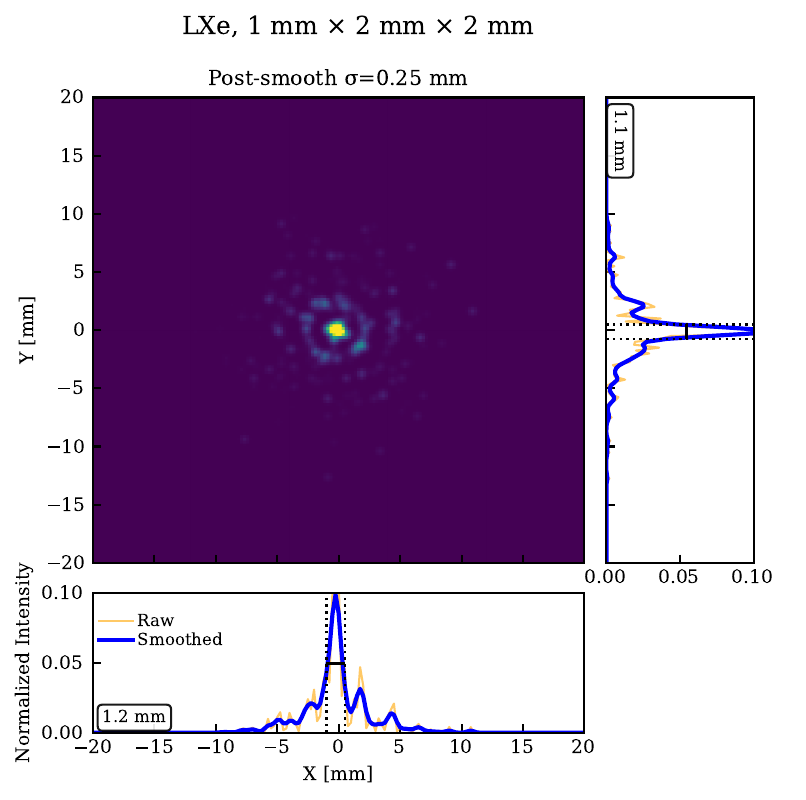}
    \caption{Reconstructed images from a simulated point-source in a single axial slice for a LYSO-based PET system (left)  and an LXe-based PET system (right). The FWHM X\&Y resolutions (numbers shown on figures) are extracted by applying a two-dimensional Gaussian fit to the smoothed reconstruction. }

    \label{fig:position}
\end{figure}

Based on these assumptions, the reconstruction results are shown in Fig.~\ref{fig:position}. LYSO-based systems exhibit reconstructed spatial resolutions of approximately 4~mm full-width-at-half-maximum (FWHM), consistent with that reported in the literature~\cite{rausch2019philipsvereosperformance}, while LXe-based systems yield spatial resolutions of approximately 1~mm FWHM. The better performance reflects the intrinsic three-dimensional position sensitivity of the LXeTPC architecture.


\section{Conclusion}
We have presented a liquid xenon time projection chamber concept for positron emission tomography and evaluated its system-level performance using Monte Carlo simulations. Compared to LYSO-based PET configurations, LXe detectors exhibit lower absolute stopping efficiency but provide improved photopeak purity and significantly enhanced reconstructed spatial resolution. These results demonstrate that LXe-based PET detectors can achieve competitive or superior imaging performance, particularly for high-resolution and scalable PET applications.

In this study, time-of-flight (TOF) information is not included in the evaluation of detection efficiency. TOF information enhances the effective sensitivity and position reconstruction of PET systems by improving localization of coincidence events along the line of response and increasing the signal-to-noise ratio in image reconstruction. State-of-the-art LYSO-based PET systems typically achieve coincidence timing resolutions of approximately 180–200~ps, limited by scintillation timing, optical photon transport, and the inability to resolve multiple interactions within a single detector element~\cite{gundacker2014time,gundacker2019highfrequencysipm}. In contrast, the fast scintillation response of liquid xenon, characterized by a prompt singlet component, sub-nanosecond effective rise time, and high photon yield~\cite{Aprile:2009dv}, places its intrinsic timing limit well below that of LYSO-based detectors~\cite{gundacker2016measurement,derenzo2014fundamental}. When combined with intrinsic depth-of-interaction reconstruction and earliest-interaction selection in multi-site events, an overall coincidence time resolution at the level of $\sim$100~ps is a realistic target for LXe-based PET, sufficient to partially compensate for its reduced stopping power.

Future work will focus on experimental validation of both detector-module and system-level performance. This includes direct measurements of energy resolution and three-dimensional position resolution in individual LXe detector modules, as well as their impact on reconstructed image quality at the system level. In parallel, integration of time-of-flight information and optimization of coincidence timing performance will be pursued. These efforts will be complemented by studies of practical implementation in medical imaging environments, including detector scalability, operational stability, and system integration.

\bibliographystyle{JHEP}
\bibliography{biblio}




\end{document}